\documentclass[conference]{IEEEtran}
\IEEEoverridecommandlockouts
\usepackage{cite}
\usepackage{amsmath,amssymb,amsfonts}
\usepackage{algorithmic}
\usepackage{graphicx}
\usepackage{textcomp}
\usepackage{xcolor}
\def\BibTeX{{\rm B\kern-.05em{\sc i\kern-.025em b}\kern-.08em
    T\kern-.1667em\lower.7ex\hbox{E}\kern-.125emX}}

\newcommand{\red}[1]{\textcolor{red}{#1}}

\newcommand{\stannis}{Stannis}
\newcommand{\laguna}{Laguna}
\newcommand{\functionname}{HyperTune}
\newcommand{\Fig}[1]{Fig.~\ref{#1}}
\usepackage{subcaption}
\usepackage{algorithmic}
\usepackage[ruled]{algorithm2e}

\newcommand{\id}[1]{\ensuremath{\mathit{#1}}}

\begin{document}

\title{{\functionname}: Dynamic Hyperparameter Tuning For Efficient Distribution of DNN Training Over Heterogeneous Systems}

\author{
 \IEEEauthorblockN{Ali HeydariGorji}
\textit{UC Irvine}\\
\and
\IEEEauthorblockN{Siavash Rezaei}
\textit{UC Irvine}\\
\vspace{4mm}
\IEEEauthorblockN{Vladimir Alves }
\IEEEauthorblockA{\textit{NGD Systems Inc.}\\}
\and
\IEEEauthorblockN{Mahdi Torabzadehkashi}
\IEEEauthorblockA{\textit{NGD Systems Inc.}\\ }
\vspace{4mm}
\IEEEauthorblockN{Pai H. Chou }
\IEEEauthorblockA{\textit{UC Irvine}\\}
\and
\IEEEauthorblockN{Hossein Bobarshad }
\IEEEauthorblockA{\textit{NGD Systems Inc.}\\}
}

\maketitle

\begin{abstract}
Distributed training is a novel approach to accelerate Deep Neural Networks (DNN) training, but common training libraries fall short of addressing the distributed cases with heterogeneous processors or the cases where the processing nodes get interrupted by other workloads. 
This paper describes distributed training of DNN on computational storage devices (CSD), which are NAND flash-based, high capacity data storage with internal processing engines. A CSD-based distributed architecture incorporates the advantages of federated learning in terms of performance scalability, resiliency, and data privacy by eliminating the unnecessary data movement between the storage device and the host processor. The paper also describes {\stannis}, a DNN training framework that improves on the shortcomings of existing distributed training frameworks by dynamically tuning the training hyperparameters in heterogeneous systems to maintain the maximum overall processing speed in term of processed images per second and energy efficiency. Experimental results on image classification training benchmarks show up to 3.1x improvement in performance and 2.45x reduction in energy consumption when using {\stannis} plus CSD compare to the generic systems.

\end{abstract}


\section{Introduction}

Nowadays, huge volumes of data are being generated continually due to the exponential growth in the number of data-capturing devices, ranging from industrial to personal IoT devices. Data gain value after being processed, and machine learning (ML) has become a mainstream way of processing the captured data.
To be effective, ML models need to be trained with a sufficiently large amount of data. The more data being used for the training phase, the more accurate the ML model can be in the inference phase. However, it often takes days or weeks to train a neural network to an acceptable accuracy level. To accelerate training, distributing the training on clusters of GPUs can be effective, although GPUs tend to be power hungry, making them unsuitable for edge devices or data centers.

In addition to performance and power, data privacy is a rising concern in a distributed system. Data encryption or approximation could help address some privacy issues \cite{encryption}, but these methods are prone to intrusion and hacking as well. Federated learning (FL) \cite{FL} is a new solution for sparse training. In FL, each device trains a local replica of the network and shares the parameter updates with a head node and other devices respectively. FL considered to be asynchronous in the way that the nodes work independently and do not wait for other nodes to finish the training. This method is not as efficient as synchronous training such as the GPU example. 
    
Our approach to addressing the above concerns is in-storage processing (ISP), which brings the processing to data rather than moving data to the processing nodes. The benefit of using ISP is that the data does not have to leave the storage system and can be processed in place. As a result, it can cross off the privacy and security concern by keeping the data within the storage system. It also reduces the movements of data, since only the processed data or the updated parameters rather than the raw data are shared with the head node. 
    
To support the ISP concept, we introduce the latest generation of Solid State Drives (SSD), called Computational Storage Devices (CSDs). CSDs are high-capacity, NAND flash-based data storage systems that are equipped with internal processing engines. Going back to the example of distributed processing, all the training procedures can be done on CSDs and only the final parameters that are significantly smaller in size than the raw input data are shared with other nodes.

To efficiently distribute the processing in a heterogeneous environment consisting of the CSDs and the host node, we examined the current distributing libraries such as Tensroflow \cite{ref_bck_tensorflow} and Pytorch \cite{ref_bck_pytorch}. Since the processing capability of CSDs are vastly different from a high end server processor such as Intel 
Xeon%
, such libraries fall short of efficiently distributing the training task. To address this problem, we introduce {\stannis}, a DNN training framework that can efficiently distributes the training process on homogeneous and heterogeneous systems. {\stannis} evaluates the processing capability of each node and assigns appropriate portion of the task to each individual node. It also monitors the training progress and reconfigures the hyperparameters to optimize the training's progress. Overall, we can summarize our contributions as follow:

\begin{itemize}
     \item Introducing {\laguna}, an M.2 form factor CSD platform,
     \item Deploying the concept of ISP for distributed ML training,
     \item Proposing a hyperparameter-tuning algorithm to optimize training operations and a load-balancing algorithm that dynamically changes training hyperparameters to mitigate the effects of the interrupting tasks and workloads,
     \item Proposing a data assignment algorithm that considers data privacy (private data is processed on CSD),
     \item Evaluating the proposed framework on a cluster of 36 CSD prototypes.
\end{itemize}

The rest of this paper is organized as follows.  Section 2 provides a background and discusses  related work on distributed ML training and ISP. Section 3 introduces the proposed distributed training method. Section 4 presents {\laguna}, our prototype CSD platform. Section 5 presents our evaluation setup and the experimental results. Finally, Section 6 concludes this paper with directions for future work.

\section{Background and related work}
\subsection{Distributed Training}
Long training times impose a severe limitation on the progress in deep neural network research. These limitations like using big dataset sizes and real-time DNN training have led researchers to emphasis on the problem of DNN training speedup \cite{iandola2016firecaffe,ioffe2015batch}. 
For inference acceleration, hardware-software co-design \cite{samragh2019codex} as well as automated DNN design algorithms \cite{javaheripi2020genecai} have been studied in the literature. \cite{samragh2019codex} propose a framework that efficiently implement a neural network on FPGA in a way that requires far less memory than the conventional implementations. This is done by using reinforcement learning to determine an optimal encoding bit-width across network layers. 

Researchers in \cite{li2014efficient} show that increasing the batch size does not necessarily decrease the convergence rate, and thus large batch sizes can be used in paralleled stochastic gradient descent training. This helps reduce the communication within the training session.

Authors in \cite{eshratifar2019jointdnn} propose an efficient, adaptive, practical engine named JointDNN for collaborative computation between a mobile device and cloud for DNNs in both inference and training phase. They provide optimization formulations at layer granularity for forward and backward propagation in DNNs. It can adapt to limits on mobile battery, constraints on cloud server load, and quality of service. 

\cite{khomenko2016accelerating} proposes an algorithm to speed up the training of the recurrent neural networks. This algorithm is based on sequence bucketing and multi-GPU data parallelization. Since the processing complexity in training is determined by the length of the data sequence in a batch, equalizing the data length helps finish the training at the same time, thus improving the speed. This method has been shown effective with Long Short-Term Memory (LSTM) and Gated Recurrent Units (GRU) training in voice recognition, language modeling, and other perceptual ML tasks.

FireCaffe \cite{iandola2016firecaffe} is an example of a framework that accelerates DNN training on a cluster of GPUs by reducing the communication overhead between the nodes. This is done by determining the best network communication hardware that fits the training session and using more efficient communication algorithms such as reduction trees instead of a generic parameter server. Another option is to increase the batch size and change other hyperparameters, which in turn decrease the number of updates and the communication cost. These methods enable FireCaffe to achieve a 36\% scaling capability for specific neural networks.

A better alternative to FireCaffe is Horovod \cite{sergeev2018horovod}, which can run on most of the current DNN libraries such as Tensorflow \cite{ref_bck_tensorflow} and pytorch \cite{ref_bck_pytorch}. It uses the NVDIA NCCL and MPI for the communication. Horovod improves the communication by using the ring allreduce algorithm \cite{ref_bck_allreduce} where each node communicates with only its two neighbor nodes in a circular fashion. It also changes the learning rate after a certain number of epochs have passed to preserve the accuracy with a better convergence rate.  By doing so, Horovod can yield up to 88\% scaling efficiency on CPUs. However, Horovod is effective on only homogeneous but not heterogeneous systems. We address Horovod's shortcomings by assigning a proper portion of processing to each node and continuously monitoring the system's performance.

\subsection{In-Storage Processing}
In-storage processing is a very promising solution to overcome the limited processing performance of the available multi-core processors by not only eliminating the data movement, but also adding extra processing power to the servers.  
We can categorize the available in-storage processing platforms into two main groups: 1) those that time multiplex the available processing units dedicated to the storage controller for the purpose of in-storage processing \cite{kim2016storage, park2016storage}, and 2) those that implant dedicated processing engines for in-storage processing \cite{torabzadeh_PDP,jun2015bluedbm, gu2016biscuit}.
While the first approach is power and cost efficient, it provides a limited processing power and also it can negatively impact the performance of the SSD controller.
In the previous works focusing on the second approach, researchers have investigated different types of dedicated processing engines such as FPGAs \cite{kim2016storage} and embedded processors \cite{Torabzadeh_bigdata , torabzadeh_hpc}. Although FPGAs are power efficient, they have several limitations for being used in ISPs, including the challenging step of implementing an RTL design of the tasks \cite{rezaei_ReConfig}, the time consuming step of reconfiguration of FPGAs for running different tasks, and the lack of supporting file systems to allow tasks to deal with the concept of file when accessing the storage.
Recently, the new concept of deploying a dedicated processor in the storage unit has been investigated and it has shown promising features. More detailed discussion about this approach is presented in Section~\ref{sec:hardware}.

\section{Method}
Most of current DNN training applications are optimized for homo\-geneous architectures such as a cluster of GPUs where all GPUs have the same processing performance and memory sizes. Applications such as Horovod go further and enable us to use heterogeneous systems such as a mixture of CPUs and GPUs for training. However, a major obstacle associated with these applications is that they use static hyperparameters for the training. It means that the training hyperparameters such as the batch size or the input dataset are determined at the beginning of the process and will remain the same throughout the entire training.
{\stannis} is an extension to the current distribution engines. \cite{heydarigorji2020stannis} It targets heterogeneous systems such as a servers with a master CPU node and clusters of CSDs or multiple servers with different CPUs. {\stannis} enables the system to monitor the training performance and tune the hyperparameters to maximize the processing speed while maintaining the accuracy. For now we use {\stannis} on Horovod and Keras, but it can be paired with other applications as well.

\subsection{Batch Size and Dataset Size Tuning}
\begin{figure}[]
  \centering
  \includegraphics[scale=0.35]{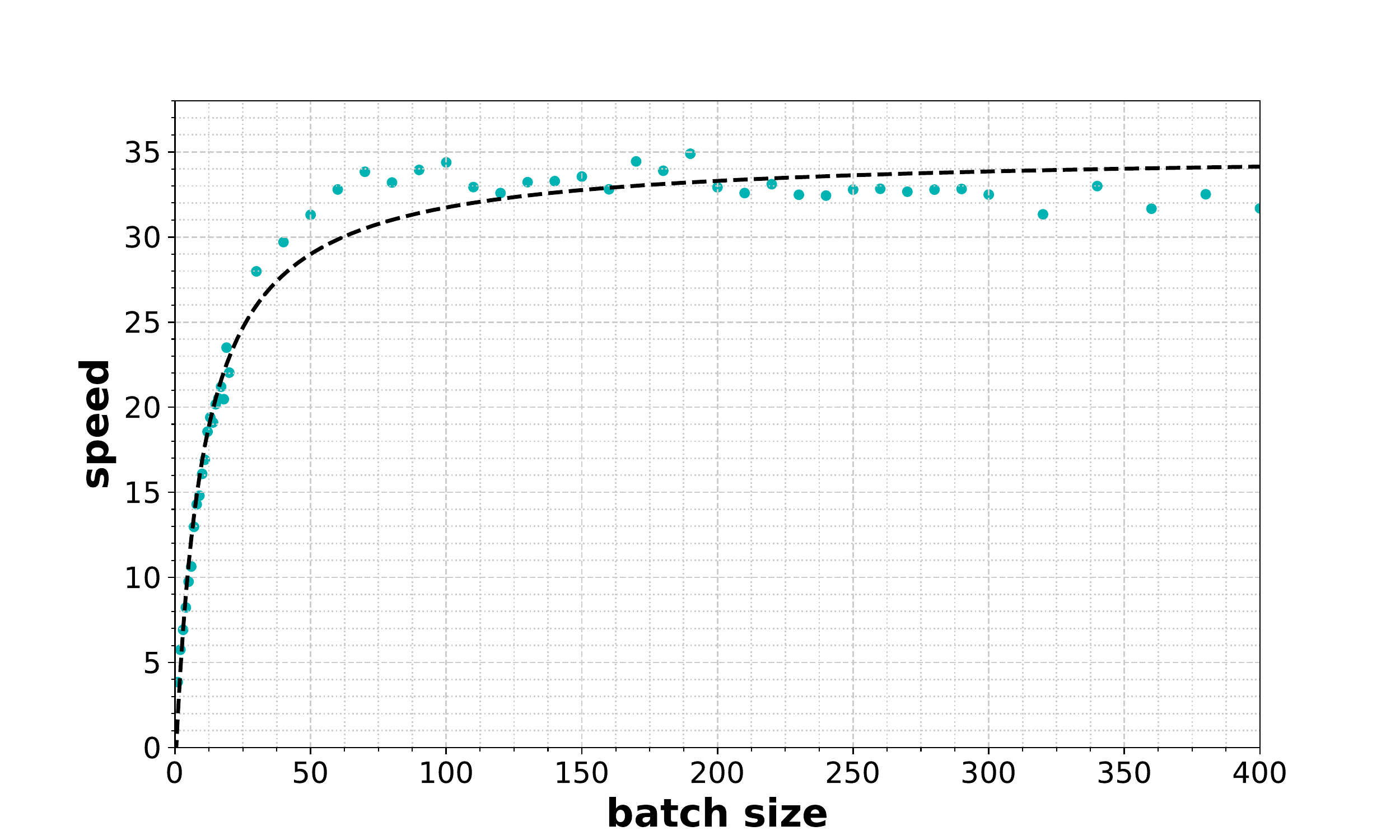}
  \caption{Processing speed (img/sec) vs batch size for MobileNetV2}
  \label{bs-sp}
\end{figure}

{\stannis} starts with benchmarking the network to be trained on all processing engines by running a small training session with different batch sizes. This procedure helps us determine the best batch size to achieve the highest processing speed on one node. It also enables us to generate a function to convert the batch size to the processing speed by curve fitting on the pairs of [batch size,speed]. \Fig{bs-sp} shows the example of such pairs for MobileNetV2. As the trend shows in \Fig{bs-sp} for the MobileNetV2, the processing speed remains the same beyond a certain batch size, which means the operation is getting more communication bound rather than computation bound. Having the speed function for all nodes, we select the most influencing  device and maximize its processing speed by choosing the proper batch size. The most influencing device is determined by multiplying a single device's processing speed by the number of such device. Having the batch size and the speed, we can calculate the required time per step. Since in the synchronized training, the synchronization happens at the end of each step, the elapsed time per step should be the same for all nodes. This helps us determine the best batch size on all other nodes such that all nodes finish at the same time, thus eliminating the idle time caused by rank stall. The portion of the dataset assigned to each node and the number of steps per epoch is calculated as follows: \\

\begin{equation}
\label{eq1}
\begin{split}
&\id{Dataset}_i= \frac{\id{BS}_i}{\sum_{i=1}^{N}BS_i}\times \id{Dataset}\\
&\id{N}_{\text{steps}}=\frac{\id{Dataset}}{\sum_{i=1}^{N}BS_i}
\end{split}
\end{equation}

where \textit{$Dataset_i$} is the portion of input data assigned to node \textit{$i$} and \textit{$BS_i$} is the batch size on node  \textit{$i$} and  \textit{$Dataset$} is the input dataset size. Assuming there are two types of data, the private and the public, {\stannis} considers the permission of the data and assigns the private data to the local CSD and the public data to the local CSD and other processing nodes. This guarantees the security of the data by eliminating the need to move the data out of the storage system. Since the input data on one node is shuffled before training, the parameters that are updated as a result of the training on the private data is mixed with parameters of the public data and thus, ensures that no private data is transmitted to outside. This is a fundamental characteristic of the Federated Learning\cite{FL}.

\subsection{Dynamic Batch and Dataset Allocation}
So far, {\stannis} provides the optimized hyperparameters for maximizing the overall speed and ensures the privacy of the data is preserved. Considering the nature of the CPU, it is possible for the processing nodes to get busy with other tasks which can affect the available processing power to the training session. Since the distributed training is a synchronous operation, an interrupt to any node means the entire operation halts until the busy node finishes its portion of the operation. Hence, it is crucial to ensure that all nodes can finish each step at the same time. {\stannis} tries to compensate for the workload interrupt in nodes by using our implemented function, called {\functionname}. {\functionname} reschedules the portion of operation that is assigned to each node based on the availability of the processing power on that node. This is done by measuring the  local processing speed and available processing power and consequently, updating the batch size list, by either decreasing the batch size on the busy node or increasing it on the other nodes. 

Since changing the batch sizes also requires a recalculation for the dataset assignment. {\stannis} reassigns the dataset based on the new batch size to prevent rank stall on the training session. New batch sizes, data set indexes and lengths are passed to each node using MPIscatter function. 

\subsection{Monitoring and Decision Making}
The initial approach was to monitor the processing speed at the end of each epoch and update the hyperparameters accordingly. This method was sub-optimal since the interruption could happen anywhere within an epoch and could finish before the end of the epoch. Hence, we decided to implement the monitoring session after each step within the epoch. The speed measurements from all nodes are gathered on the 
one arbitrary node using MPIgather and passed to a decision making function. 

To make a better decision, the speed change along with the progress percentage of the current epoch are passed to the function and converted to a decline index based on the following weighted sum:

\begin{equation}
\label{eq2}
\id{index_i}= \id{0.7}\times\frac{\id{SP}-\id{SP_i}}{\id{SP}} + \id{0.3}\times(\frac{\id{N_{step}}-\id{step_i}}{N_{step}})
\end{equation}

\textit{$SP$} is the normal speed that is obtained form the \textit{$batch size\_to\_speed()$} function. \textit{$SP_i$} and \textit{$step_i$} are the current speed and step. and \textit{$N$} is the number of steps per epoch. To avoid chattering, a hysteresis-like algorithm is implemented to ignore glitches and mis-measurements in speed. When the decline index exceeds 20\%, the step will be flagged as under-utilized and this report will be saved in a separate array.  Having 5 consecutive under-utilization terminates the current epoch and triggers the \textit{$batch size\_controller()$} function to determine a new batch size.
Our initial approach to determine the new batch size was to use the inverse of the \textit{$batch size\_to\_speed()$} function 
we got from the tuning section and determine the new batch size based on the current processing speed of the interrupted node. Although this method sounds promising, our evaluations showed non-negligible error that could worsen the performance. As a second solution, we decided to use a weighted averaging method from the nearest values to the current speed.  We use the following equation to calculate the new batch size:
\begin{equation}
\label{eq2}
\id{BS_i=BS_n \times \frac{SP_i - SP_n}{SP_{n+1}-SP_n} + BS_{n+1} \times \frac{SP_{n+1} - SP_i}{SP_{n+1}-SP_n}} 
\end{equation}
where $BS_i$ is the new optimal batch size and $SP_i$ is the current speed that is between $SP_{n}$ and $SP_{n+1}$. A third method that yields almost the same result is to monitor the CPU usage of the training session on each node. For this approach, we have implemented a sliding window to keep track of the CPU usage for the last 10 steps. The new batch size is proportional to the  average of the last five steps with the declined CPU usage and the normal CPU usage. Note that parameters such as the size of the sliding window or the margin for speed decline detection can be changed based on the required precision.
A benefit of using CPU utilization as a gauge
is that the system can also increase the batch size in case there is extra processing power available. When the system frees the occupied processing power, the training session can claim it back by increasing the batch size. Whereas the case of speed monitoring cannot detect the availability of extra processing power.

By rapidly monitoring the performance and updating the hyperparameters, {\stannis} assigns a larger portion of processing to the free nodes to mitigate the overall performance impact of interruption. One concern is that by early termination of epoch, the network will lose part of the  dataset in the training process and multiple occurrences of such epoch terminations might lead us to completely miss that portion of the dataset. The solution to this concern is to use the shuffle function when creating the input batch of the data. This results in the randomness of the input data list and statistically ensures that all input data will go through the training after a sufficient number of epochs. 
Another concern is that the training might not converge as a result of constant changing of batch size. To address this concern, we change the batch size in a limited range such that it will not affect the convergence. As an option, we can change the learning rate along with the batch size to ensure a better convergence rate and higher achieved accuracy, This method is currently not implemented but will be added to the decision making function in the future updates.


\section{Computational Storage Platform}
\label{sec:hardware}
In this section, we introduce {\laguna}, a full-featured ASIC-based CSD that enables us to run the ML training experiments in a realistic environment. 
The proposed and prototyped CSD interacts with the host via the NVMe over PCIe protocol and has 8~TB of flash storage.
We first describe the hardware and software architecture of {\laguna} CSD, including different components and how they communicate with each other. We also demonstrate a fully functional {\laguna} prototype and provide implementation details and features of the CSD.

\subsection{{\laguna} Hardware Architecture}

\begin{figure}
\centerline{\includegraphics[scale=0.5]{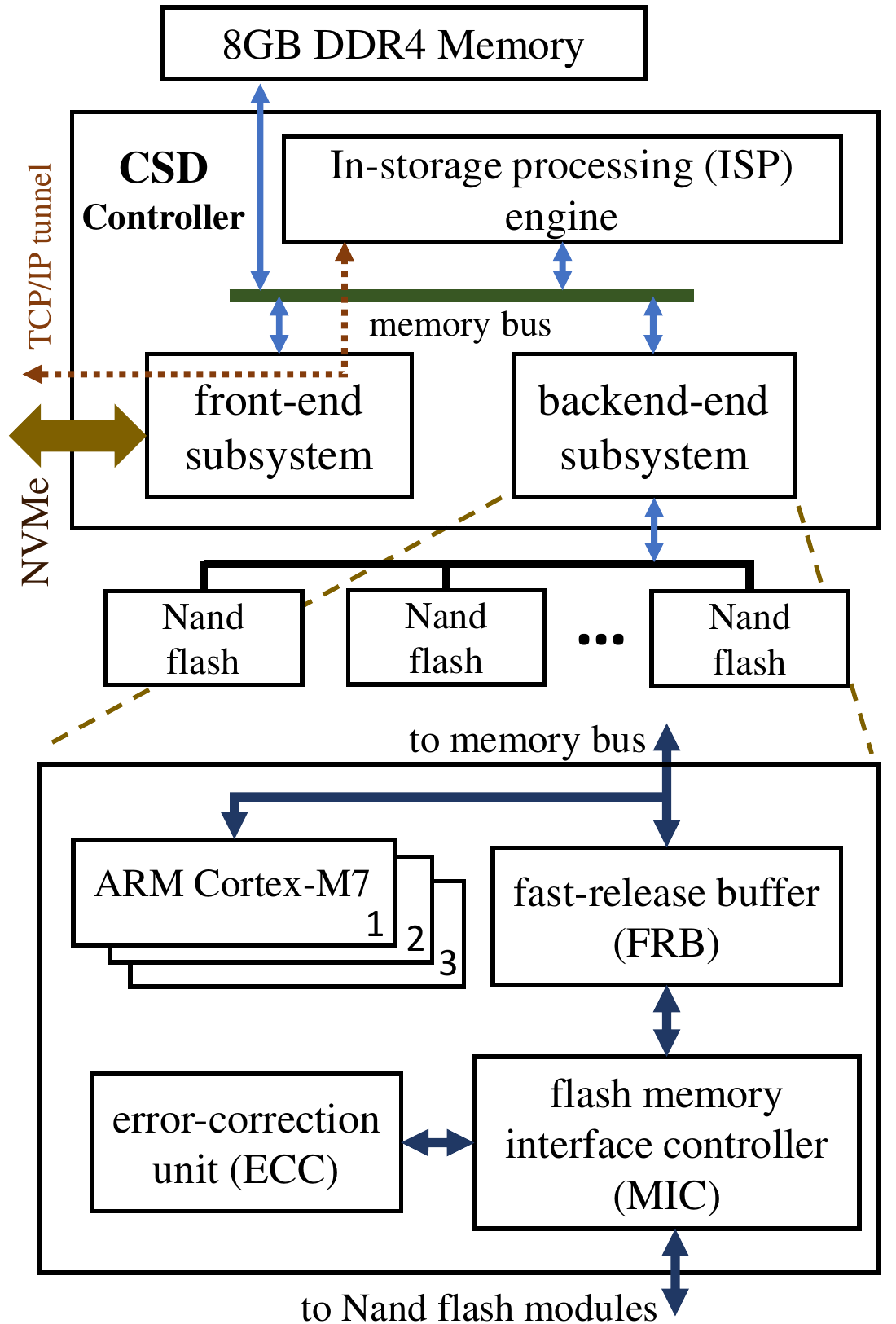}}
\caption{{\laguna}'s high-level architecture}
\label{lug_arch}
\end{figure}

\Fig{lug_arch} shows the high-level architecture of {\laguna} CSD. The proposed CSD is composed of three main components: front-end, back-end, and in-storage processing (ISP)  subsystems. The front-end (FE) and back-end (BE) subsystems are similar to the subsystems in the off-the-shelf flash-based storage devices, i.e., SSDs.  In {\laguna}, a conventional subsystem runs the flash translation layer (FTL) processes that manage flash memory and transfer data between the host and NAND flash modules.
Besides this conventional subsystem, {\laguna} also has a complete ISP engine inside, which runs user applications in-place without sending data to the host.
While ISP technology is simple in definition and concept, designing and implementing a practical CSD platform is complicated due to the coexistence and interaction of the conventional components and the innovative ISP engine.

\subsubsection{Back-End}

For managing the NAND flash modules, BE has three ARM Cortex-M7 processing cores together with a fast-released buffer (FRB), an error-correction unit (ECC), and a memory controller interface unit (MIC).  
The ARM Cortex M-7 cores run the flash translation layer (FTL) processes, including logical and physical address translation, garbage collection, and wear-leveling. The ECC unit is responsible for handling the errors that happen in the flash memory modules. The fast-release buffer (FRB) transfers data between the memory interface controller (MIC) and other components, while MIC issues low-level I/O commands to NAND flash modules. These modules are organized in 16 channels, so 16 I/O transfers can be performed simultaneously.

\subsubsection{Front-End}

The FE is responsible for communicating with the host via the NVMe protocol. It receives the I/O commands from the hosts, interprets them, checks the integrity of the commands, and populates internal registers to notify the BE that a new I/O command is received. The FE also de/packetizes the data that is transferred between the host and the CSD. This subsystem consists of a single-core ARM Cortex-M7 processor and an ASIC-based NVMe/PCIe interface.

\subsubsection{ISP Engine}
Besides the FE and BE subsystems, there is an ISP engine inside {\laguna}, which is composed of a quad-core ARM Cortex-A53 processor together with a software stack that provides a seamless environment for running user applications. The ISP engine has access to the shared 4~GB memory. A full-fledged Linux OS has been ported to the ISP engine, so the ISP engine supports a vast spectrum of programming languages and models. The ISP engine has a low-latency, power-efficient direct link to the BE subsystem to read and write the flash memory. In other words, the data transferred to the ISP engine bypasses the whole FE subsystem and the power-hungry NVMe/PCIe interface. In Section~\ref{sec:exp_res}, we show how this data link together with the efficient ISP processing subsystem and the {\laguna} software stack benefits the ML applications.

\begin{figure}
\centerline{\includegraphics[scale=0.5]{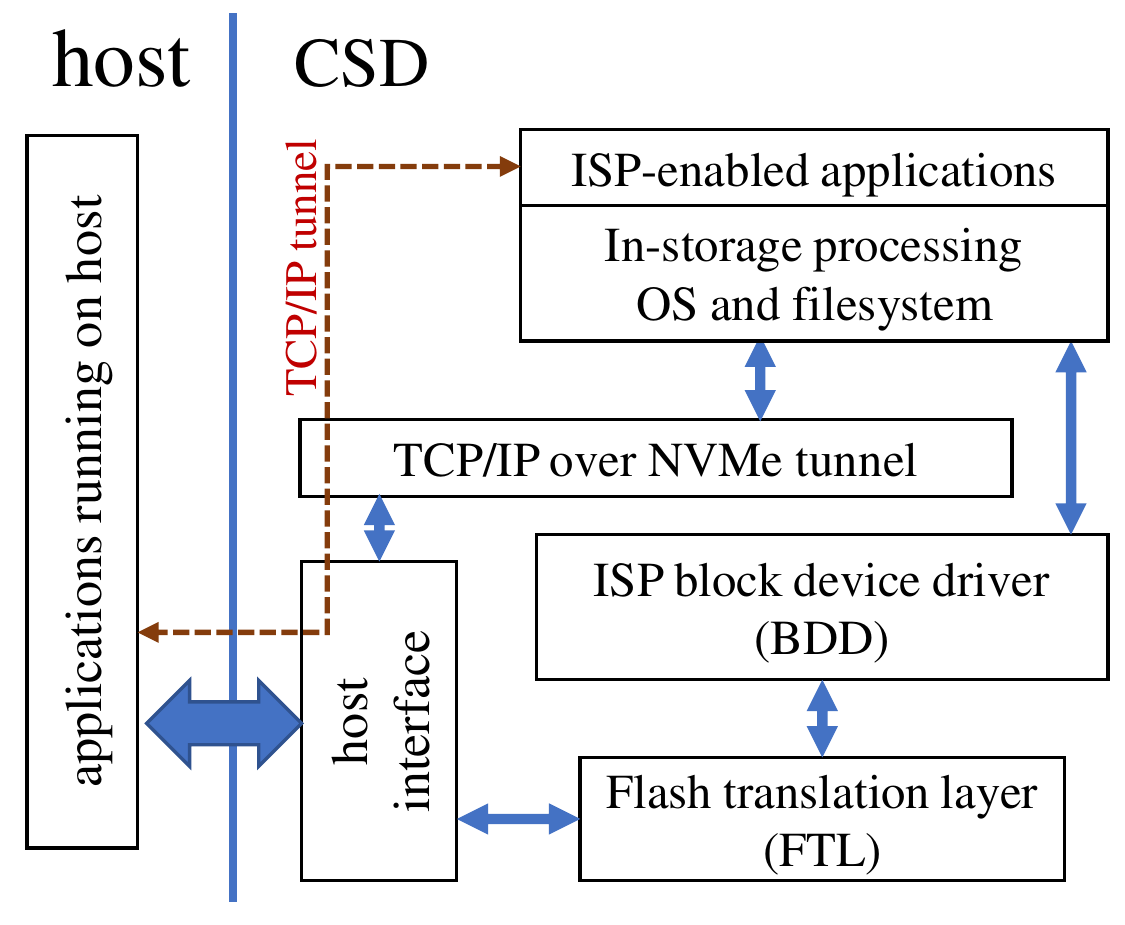}}
\caption{{\laguna}'s software architecture}
\label{soft_arch}
\end{figure}

\subsection{{\laguna} Software Architecture}

\Fig{soft_arch} depicts the {\laguna} software architecture that enables running user applications in-place.
From the ISP user's point of view, it is crucial to have a standard communication link between the host and the ISP engine. The user on the host side can use this communication link to initiate the execution of the application inside the ISP engine and monitor the outcome of the executions. To provide such a standard link, a TCP/IP tunnel through the NVMe/PCIe is supplied inside the {\laguna} ISP engine. In other words, host-side applications can communicate to the applications running inside the CSD via a TCP/IP link. This link is also essential for distributed processing applications where processes running on multiple nodes require a TCP/IP link to communicate. To implement this link, we have developed a software layer on both {\laguna} and the host to de/packetize the TCP/IP data inside the NVMe packets. On the {\laguna} side, the TCP/IP tunnel software layer runs on the FE subsystem.

On the other hand, the ISP applications should be able to read data from the flash memory in a power-efficient and low-latency manner. The ISP block-device driver (BDD) module is developed as an interface between the BE subsystem and the applications running in the ISP engine. This BDD module makes it possible to mount the flash media in the ISP operation system, so the ISP applications can read/write data from/to flash media similar to when they run on a conventional machine. In other words, the ISP hardware and software architecture is entirely invisible to the applications, so the ISP users do not need to modify the applications to offload them to {\laguna}. Since both the OS running on the ISP engine and the host’s OS can access the flash memory simultaneously, we have used the Oracle cluster filesystem second version (OCFS2) for synchronization between these two operating systems. This approach takes advantage of the TCP/IP tunnel to send the synchronization control data and enforce data integrity.

\subsection{Prototype and Features}

\begin{figure}
\centerline{\includegraphics[scale=0.55]{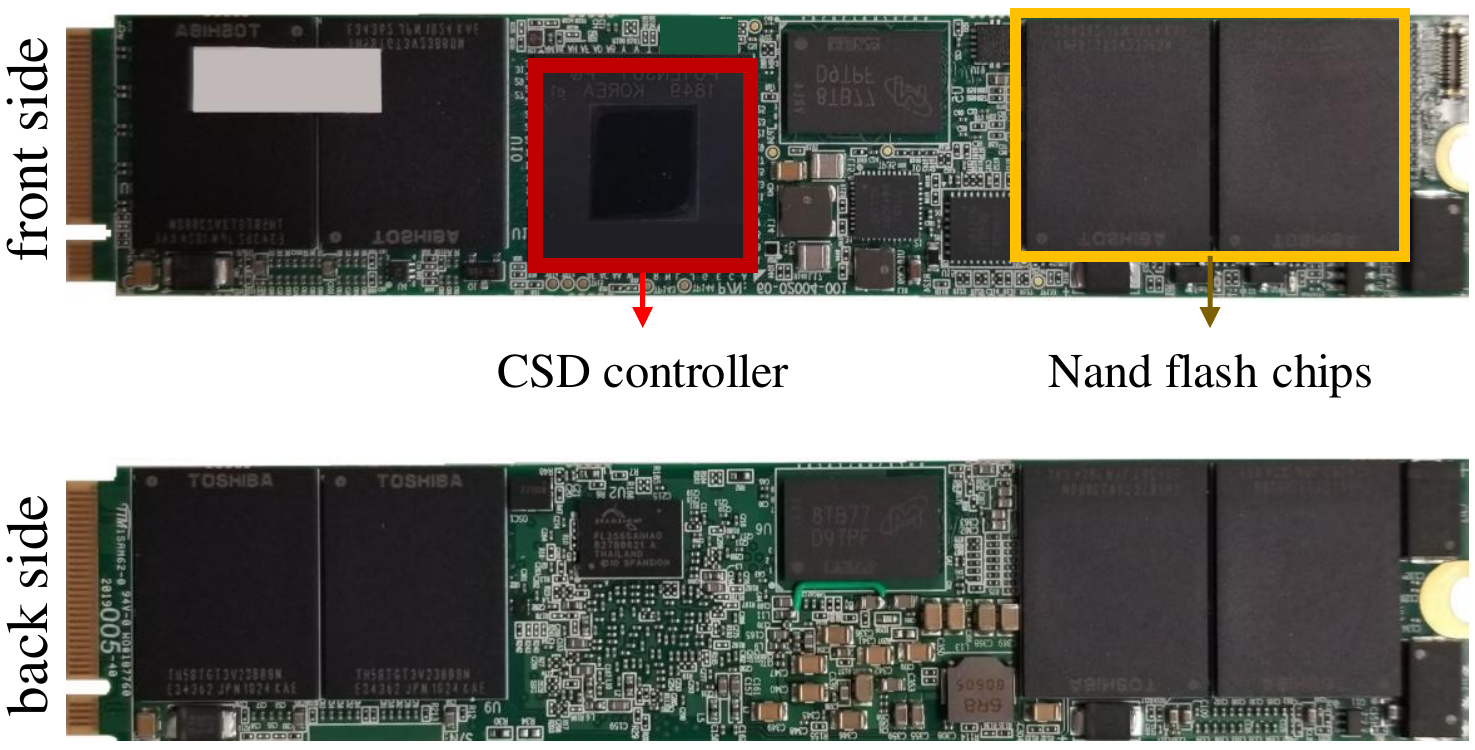}}
\caption{{\laguna} CSD prototype}
\label{lagu_proto}
\end{figure}

To demonstrate the feasibility of the {\laguna} design, we developed a fully functional prototype, as shown in Fig.\ \ref{lagu_proto}. As previously described, the prototype is an NVMe over PCIe CSD. The categorized features of the CSD comes in Table \ref{lagu_feat}.

\begin{table}
\centerline{\includegraphics[scale=0.47]{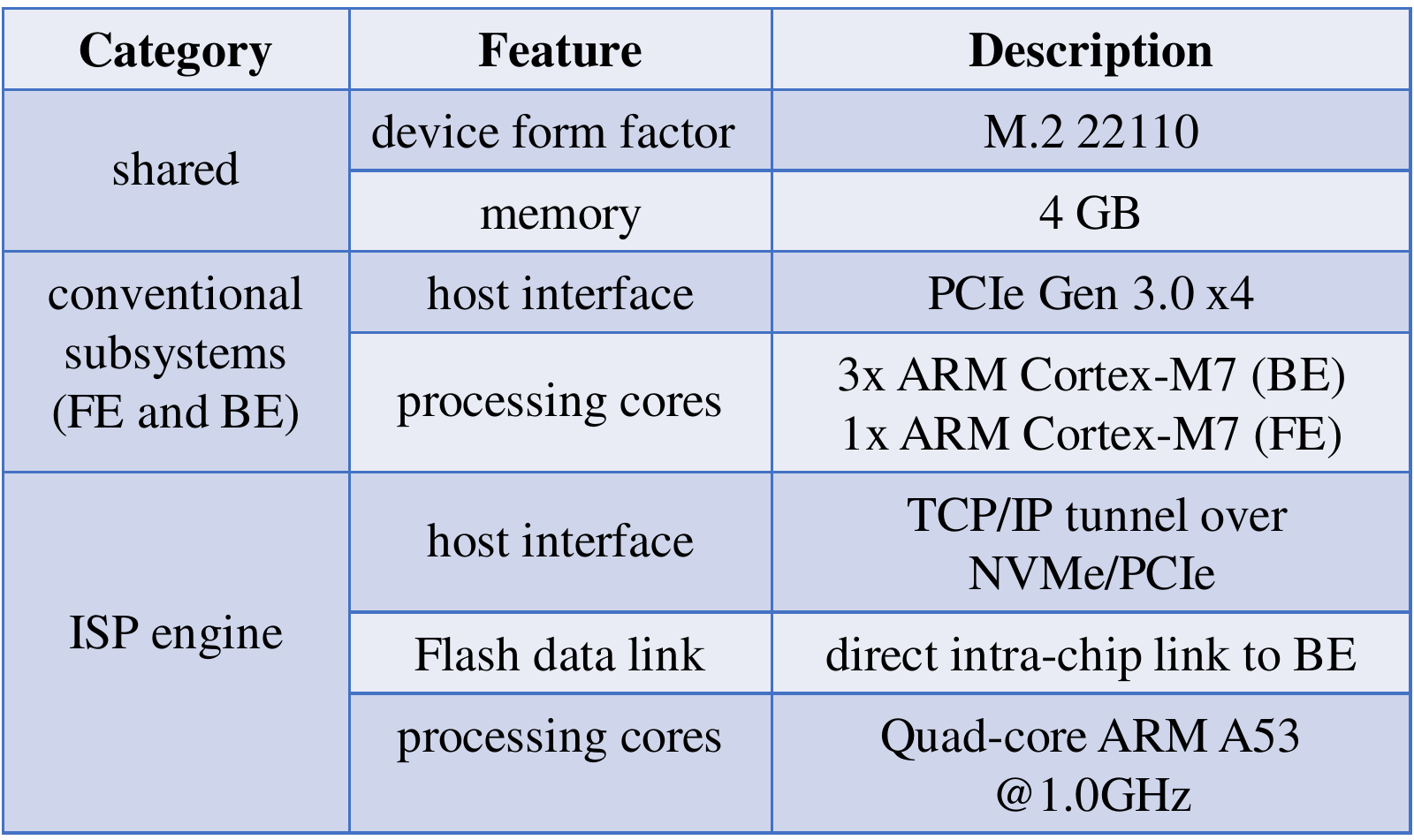}}
\caption{{\laguna}'s prototype features}
\label{lagu_feat}
\end{table}

To the best of our knowledge, {\laguna} is the first fully functional M.2 form factor, ASIC-based CSD prototype that can run a vast spectrum of user applications in-place. This CSD provides a complete software stack that makes the ISP implementation details invisible to the user, so the applications can be offloaded to run in {\laguna} CSD without modification. The user can initiate and monitor the execution of the ISP applications using the TCP/IP tunnel. Additionally, multiple {\laguna} CSDs can be orchestrated in a distributed processing platform. They can join the host machine in such a distributed environment and seamlessly augment their efficient processing horsepower to the host. Later in this paper, we use this feature to run ML training on the host and multiple {\laguna} CSDs in a distributed fashion.

\section{Experimental results}
\label{sec:exp_res}
In this section, we present two experiments to evaluate different features of the {\stannis} and {\laguna}. In the first experiment, we evaluate the performance of the {\stannis} on three similar nodes in term of processing capabilities. In the second experiment, we put both {\stannis} and CSDs in use and run multiple training sessions with different configurations to investigate the performance improvement in different situations.

\subsection{{\stannis} Evaluation}
To evaluate {\functionname}, We run a training session on three similar nodes in the term of processing power. The reason for choosing similar processing power is that the significance of {\functionname} can be shown better when the nodes are equally important in the term of processing power. In this test, each node is an AIC 2U-FB201-LX server and is equipped with an 8-core, 16-thread Intel\textsuperscript{\textregistered} Xeon\textsuperscript{\textregistered} Silver 4108 CPU with 64GB~GB of DRAM. We choose the MobileNetV2 neural network to run the test on, with 300,000 images as the input for the system. To simulate external workloads, we use the Gzip compression application which enables us to occupy the desired number of cores on the processors. We decided to interrupt one node at a time to make the evaluation less complicated. In theory, interrupting multiple nodes is not different from interrupting a single node since the master node collects the interrupt reports from all nodes and can easily detect if the slowdown is due to the local workload or an external node. 

\begin{figure*}[t]
  \centering
  \includegraphics[width=\textwidth]{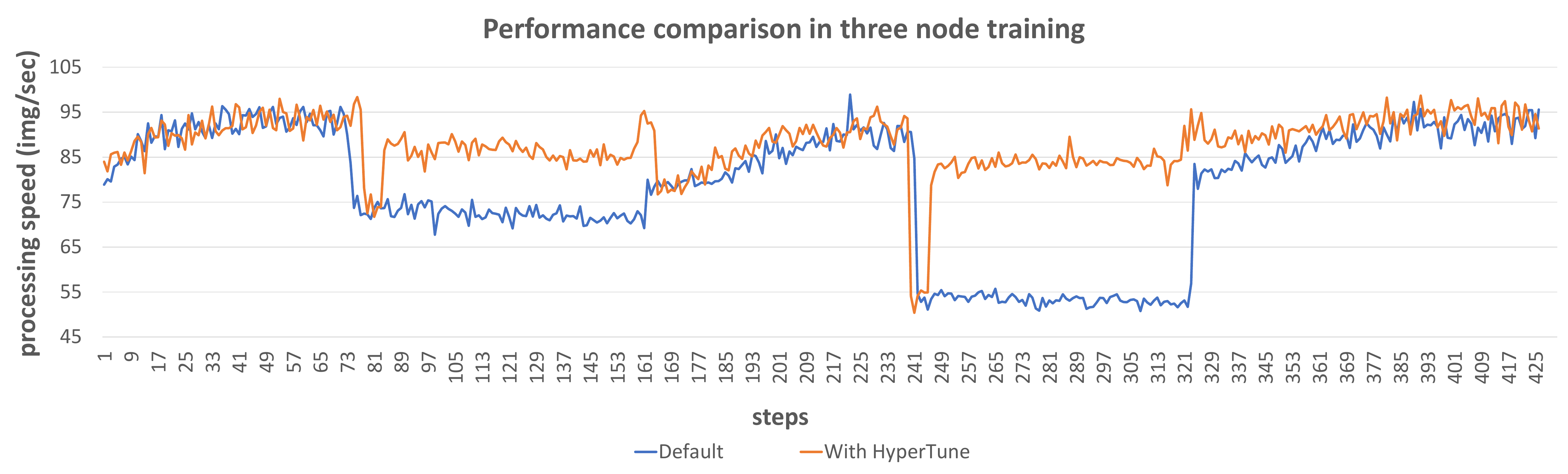}
  \caption{Evaluation of {\functionname}}
  \label{three-node}
\end{figure*}

{\stannis} starts with finding 
the best batch size for each node which is 180. Then, in two separate steps on separate nodes, we set the Gzip to occupy 4 and 6 cores out of 8 cores and monitor the speed change.  We define the performance as the average number of processed images per second for the training. \Fig{three-node} shows the overall processing speed of the three nodes during the test. As can be seen, the total processing speed is 93.4 images/second in normal operation mode. As the workload increases, the speed drops to 75.6 img/sec for 50\% external workload 
and 53.3 img/sec for 75\% external workloads. These speeds remain nearly constant as long as the workload sustains thereafter. In contrast, if {\functionname} detects interrupts by new workload and determines non-transient decline in the speed, it recomputes the new batch size for the busy node and updates the hyperparameters, which are 140 and 100 for 4 cores and 6 cores workloads, respectively. By changing the batch size, the other two nodes regain the initial processing speed which shows that the equation \ref{eq2} works fine in determining the new batch size. The processing speed declines to 83.7 and 85.8 img/sec for 75\% external workload and 50\% external workload. That is, {\functionname} is 14\% and 57\% faster than the baseline at no cost of power or accuracy and with negligible computation. 
\begin{figure}[]
  \centering
  \includegraphics[width=\columnwidth]{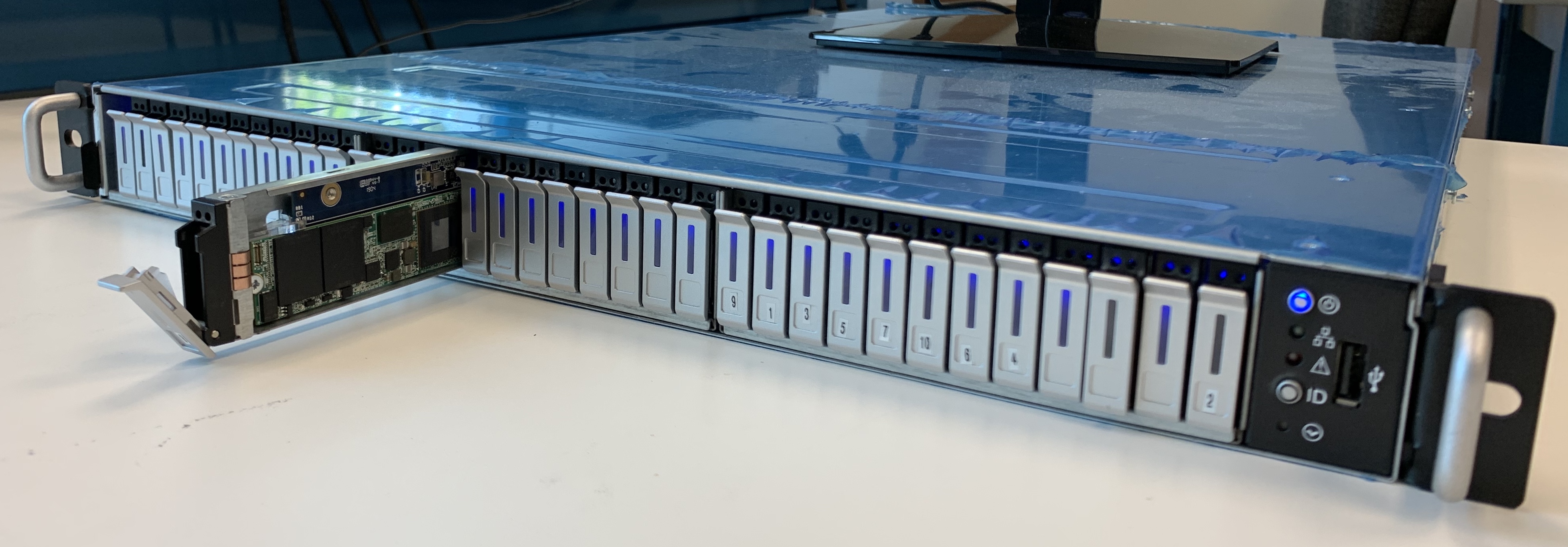}
  \caption{The 1U server equipped with 36 CSDs}
  \label{server}
\end{figure}

\subsection{{\stannis} and CSD Evaluation}
To evaluate {\stannis} on CSD, we run a training session on a custom-built FlacheSAN1N36M-UN server equipped with the same Intel
Xeon 
Silver 4108 CPU and 64~GB of DRAM. We put 36 {\laguna} CSDs, each with 8~TB capacity on the server to make a 1U class server with 288~TB storage capacity, as shown in \Fig{server}. To measure the power and energy consumption, we connect an HPM-100A power meter between the server and the power source and measure the total power consumed by the server, including the cooling system and the storage devices. The power meter logs power at 1~Hz. Since the performance and the computation load depend on the network type and size, we run our test for two different networks of different sizes, MobilenetV2 \cite{mobilenet} and ShuffleNet \cite{shufflenet}. The former is a neural network for image classification, feature extraction, and segmentation with 3.4~M parameters and 300~M multiply-accumulate (MAC) operations. The latter is a high-accuracy neural network with 5.4~M parameters and 524~M MACs, where the relatively low number of FLOPS makes it suitable for edge devices.

\begin{figure}[h]
\begin{subfigure}{.5\textwidth}
  \centering
  \includegraphics[scale=0.6]{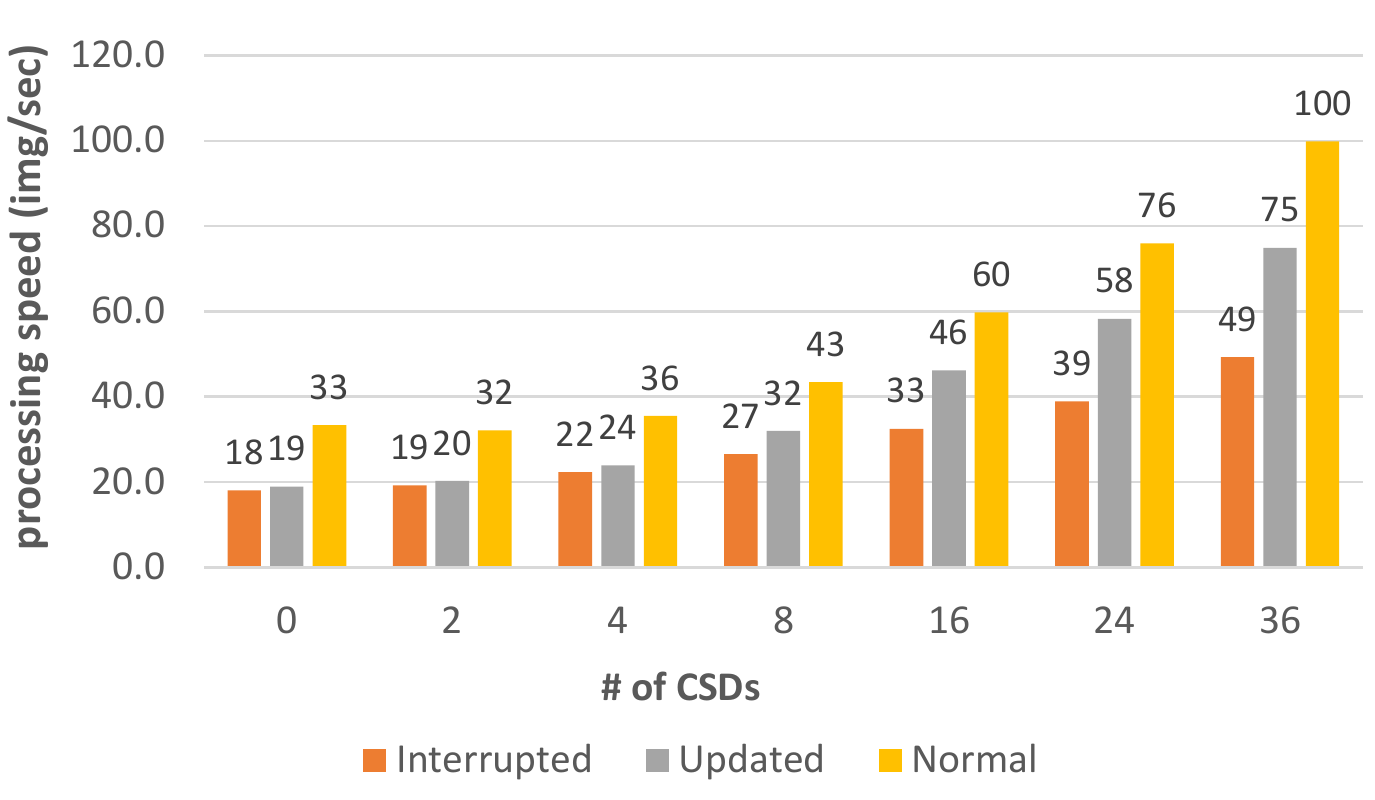}
  \caption{MobileNetV2}
  \label{fig:mobilenet}
\end{subfigure}
\begin{subfigure}{.5\textwidth}
  \centering
  \includegraphics[scale=0.6]{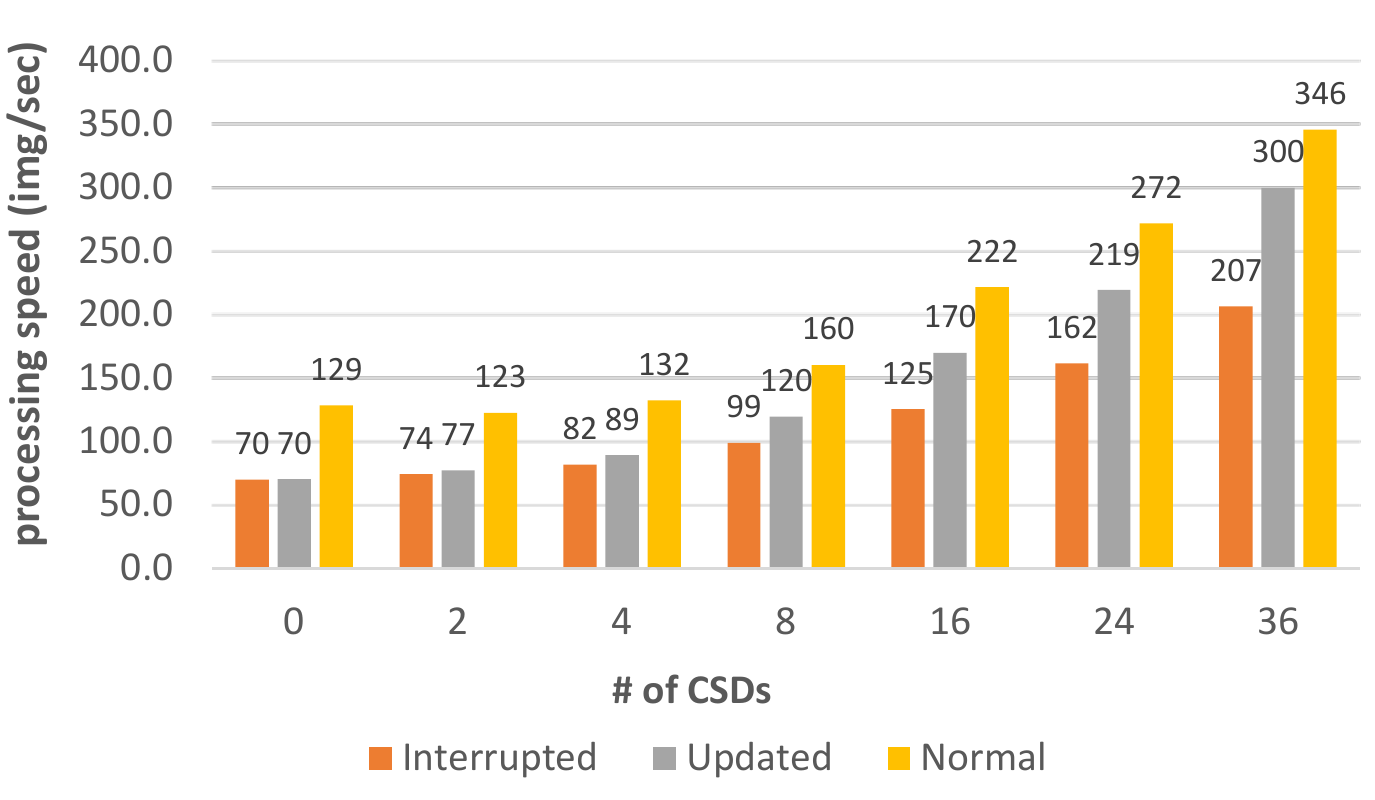}
  \caption{ShuffleNet}
  \label{fig:shufflenet}
\end{subfigure}
\caption{Benchmark results}
\label{fig:fig}
\end{figure}

Just like the previous experiment, we run multiple training sessions with different numbers of CSDs and interrupt the master node by taking 6 out of 8 cores from the training session. We only interrupt the host processor since the CSDs provide an augmented processing power and are not counted as the main processor. The initial benchmarking results give the batch size of 180 and 15 for the host and {\laguna} CSDs for MobilenetV2 and 300 and 25 for the ShuffleNet. \Fig{fig:mobilenet} and \Fig{fig:shufflenet} show the processing speed based on the number of cards in default operation mode, with external workload when {\functionname} is not engaged, and when {\functionname} is enabled. The results show that for MobileNet training, the processing speed improves from 33.4 img/sec to 99.83 img/sec when the training is distributed over the host and 36 CSDs which is 3.1x improvement. When the external workloads occupy 6 out of 8 cores, the processing speed declines from 99.83 img/sec to 49.26 img/sec. However, by engaging {\functionname}, the processing speed declines from 99.83 img/sec to 74.89 img/sec. This shows that the {\functionname} improved the processing speed about 1.5x while having no significant effect on the training accuracy and loss numbers or the power consumption. Same analysis for ShuffleNet gives us 2.82x and 1.45x performance improvement. 

We also compare the energy consumption with and without the CSD. We estimated the overall energy consumption in terms of joules per image (J/img) by integrating the power consumption over time for the entire epoch and divide it by the number of processed images per epoch. For the MobileNetV2 the energy consumption is  1.32 J/img for the host only vs.\ 0.54 J/img with 36 CSDs added to the process. It means that the entire system consumes 2.45x less energy by training on the CSDs compared to a standalone CPU.

\section{Conclusions}
In this paper, we introduced a hardware and software combination to tackle the problems associated with the distributed training of neural networks over heterogeneous systems. Our hardware named {\laguna}, a NAND flash based computational storage device, can accelerate DNN training while reducing energy consumption with the privacy and security benefits of federated learning. However, much of the potential gains in energy and performance are diminished in straightforward schemes especially on heterogeneous nodes unless optimized dynamically. To address this challenge, our software framework named {\stannis} optimizes the training hyperparameters dynamically by monitoring workload variations. {\stannis} is complementary to current DNN training engines by optimizing training at the distribution level. Our experimental results show that {\stannis} along with {\laguna} can increase the processing speed by up to 3.1x while reduces the energy consumption by 2.45x at no cost or loss of accuracy. For future work, we plan to extend hyperparameter tuning to include learning rate to increase the accuracy while providing more support for federated learning.

\bibliographystyle{IEEEtran}
\bibliography{main}

\end{document}